\documentstyle[12pt,preprint,oldlfont,aps]{revtex}
\begin{document}
\preprint{SLAC-PUB-7166, U.C. Irvine TR-96-12}
\title{
Anomalous Chromomagnetic Moments of Quarks \\
and Large Transverse Energy Jets}
\author{Dennis Silverman}
\address{Department of Physics and Astronomy \\
University of California, Irvine \\
Irvine, CA 92717-4575 }
\date{\today}
\maketitle

\begin{abstract}
We consider the jet cross sections for gluons coupling to quarks with
an anomalous chromomagnetic moment.  We then apply this to the
deviation and bounds from QCD found in the CDF and D0 Fermilab data,
respectively, to find a range of possible values for the anomalous
moments.  The quadratic and quartic terms in the anomalous moments can
fit to the rise of a deviation with transverse energy.  Since previous
analyses have been done on the top quark total cross section, here we
assume the same moment on all quarks except the top and find the range
$|\kappa'| \equiv |\kappa/(2 m_q)| = 1.0\pm 0.3$ TeV$^{-1}$ for the
CDF data.  Assuming the anomalous moment is present only on a charm or
bottom quark which is pair produced results in a range $|\kappa'_{b,c}| =
3.5 \pm 1.0 $ TeV$^{-1}$.  The magnitudes here are compared with
anomalous magnetic moments that could account for $R_b$ and found to
be in the same general range, as well as not inconsistent with LEP and
SLD bounds on $\Delta \Gamma_{\text{had}}$.

\end{abstract}

\section{Introduction}
There are several higher dimensional operators that can be used to
evaluate structure from a theory beyond the standard
model\cite{peskin}.  In this paper we use the time-honored anomalous
magnetic moment as the indicator
\cite{brodskydrell,shawsilvermanslansky,shawsilverman,rizzo1}, applied
to the quark-gluon vertex, or the anomalous chromomagnetic moment.
Several papers have been written analyzing the contribution of such a
moment to the top production cross
section\cite{soni,rizzocs,rizzolep,haberl,cheung}.  In this paper, we
assume such a moment could apply to any or all of the mainly
annihilated, scattered or produced quarks other than the top quark.
The results of this will be different from those of the four-Fermion
contact interaction operators, and should also be included in
evaluations of possible structure.  Along with the exchange of a spin
3/2 excited quark state\cite{bander}, the $E_T$ dependence and angular
two-jet behavior could help further limit or establish a composite
mechanism, or new interactions from higher mass scales.  The same
theory that gives an anomalous chromomagnetic moment may also give a
four-Fermion contact interaction, but here we only analyse the
anomalous moment mechanism in a general context.  For a complete
analysis of all color current contributions in a particular model, we
refer to the recent analysis of the Minimal Supersymmetric Standard
Model\cite{ellis}.  The key point in our analysis is that the
anomalous moment term $\kappa' \sigma_{\mu\nu} q^\nu$, when compared
to the Dirac current $\gamma_\mu$, grows at high $E_T$ as ${\cal
  O}(\kappa' E_T)$.

In the top production papers, gluon fusion or quark fusion to a
virtual gluon are calculated to produce the $t-\bar{t}$ pair.  Here we
include all quark gluon hadronic processes since quarks commonly
present in protons and antiprotons could have the small anomalous moments.

Since next-to-leading order QCD corrections have not been calculated
for this general set of anomalous moment processes, to compare with
the data we follow the CDF\cite{CDF,CDFb} procedure of calculating the
ratio of the theory with structure divided by lowest order QCD and
comparing it with the ratio of data divided by NLO QCD.  We find that
an anomalous chromomagnetic coefficient $|\kappa'|=|\kappa/(2
m_q)|=1.0$ TeV$^{-1}$ fits the CDF rise, and would be roughly upper
bounded by $1.3$ TeV$^{-1}$, and lower bounded by $0.7$ TeV$^{-1}$.
The central value for the rise in CDF is not directly supported by the
D0\cite{D0} data, but is within the one sigma systematic energy
calibration error curve for D0, which rises to 120\% at $E_T = 450$
GeV.  

We use $\kappa'$ since in the general case the internal diagram or
dynamics might not involve the light external quark, and a new physics
model calculation will give $\kappa'$ directly.  The use of the
breakup into a vector current ($\gamma_\mu$) and an anomalous
chromomagnetic moment term ($i\kappa'\sigma_{\mu \nu}$) includes all
anomalous moment vertex corrections in $\kappa'$ including those of
QCD.  However, at the very large momentum transfers we are considering
here, there are form factors on the QCD vertex corrections making up
the anomalous chromomagnetic moment either for virtual gluons or for
high $E_T$ virtual quarks (``sidewise form factors''\cite{thesis})
which will damp like ($\kappa'_{\text{QCD}} \approx {\cal
  O}(m_q/p^2_\perp)$) and become irrelevant.  At some $q^2$ the
anomalous moment from new or composite interactions will also evidence
a form factor.  That is automatically included in the analysis by
considering $\kappa'(q^2)$ a function of $q^2$, but in the comparison
to the data we do not need to invoke that dependence yet.  We test
whether an anomalous magnetic moment equal to the anomalous
chromomagnetic moment possibly indicated here would be in conflict
with the $\Gamma_{\text{had}}$ accuracy at LEP, and find it would not
be.

Due to discrepancies in the total hadronic cross sections for charm
and bottom production at LEP and SLD, $R_c$ and $R_b$, we also find a
separate range for either the charm or bottom quark only having an
anomalous chromomagnetic moment, using the quark-antiquark production
cross sections.  The range for the CDF data is $|\kappa'_{b,c}| = 3.5 \pm 1$
TeV$^{-1}$.  We note that if the $R_b$ discrepancy is accounted for by
an anomalous magnetic moment, it is the same order of magnitude as the
anomalous chromomagnetic moment found here.  $A^b_{\text{FB}}$ is not
inconsistent with this interpretation and rules out one of two
possible anomalous magnetic moment values.  For comparison with
results from the top quark total cross section from
Ref.~\cite{rizzolep}, we note from their Fig. 3, using present CDF and
D0 data on $\sigma_{\text{top}}$, that $0 \leq \kappa^g_t \leq 0.35$.
This corresponds to $0 \leq \kappa^{\prime g}_t \leq 1$ TeV$^{-1}$,
and is better than the inclusive jet limits for the $b$ or $c$ quark
alone calculated here.

\section{Anomalous Chromomagnetic Moment Cross Sections}
The top gluon fusion production\cite{rizzocs,haberl,cheung} with
anomalous chromomagnetic moments has already been calculated.  We use
the simplifications of these for the case of zero quark mass in the
various channels in which quark-quark-gluon-gluon processes occur.
For completeness we repeat these formulas here for the massless case
along with the massless QCD contribution.  We also present here the
result with anomalous chromomagnetic moments for quark - antiquark
annihilation and scattering to the same quark - antiquark with
interference, and its crossed quark-quark scattering.  We then present
the cross sections for different quarks scattering and annihilating.
The cross sections are even in powers of $\kappa'$ due to the zero
quark mass limit being used.  This is because at zero mass only gamma
matrices occur in the QCD trace and substitution of an anomalous
moment $\sigma_{\mu \nu}$ term for a $\gamma_\mu$ would give an odd
number of gamma matrices and resulting in a zero trace, unless
accompanied by an additional $\sigma_{\mu \nu}$ term.  In other terms,
at zero quark mass, there is no interference between helicity
conserving and helicity flip processes: helicity conserving ones have
either no anomalous moment terms or two such terms; helicity flip ones
have one anomalous moment for each helicity flipped quark line, and
this gets squared in the cross section.  Neglect of the mass in a
propagator term could give an error of order $m_b/E_T \approx 2.3 \%$
in the amplitude at $E_T = 200$ GeV, or $4.5\%$ in the cross section.

\subsection{Gluon Processes}
The well known result\cite{EHLQ} for gluon-gluon scattering is
\begin{equation}
\frac{d\hat{\sigma}}{d\hat{t}} = 
\frac{\pi \alpha_s^2}{\hat{s}^2} \frac{9}{2} \left(3 - 
\frac{\hat{t} \hat{u}}{\hat{s}^2} - 
\frac{\hat{u} \hat{s}}{\hat{t}^2}
-\frac{\hat{s} \hat{t}}{\hat{u}^2} \right)
\end{equation}

For quark-antiquark to gluon-gluon\cite{combridge,EHLQ,rizzocs,haberl}
the cross section is
\begin{equation} 
\frac{d\hat{\sigma}}{d\hat{t}} = 
\frac{\pi \alpha_s^2}{\hat{s}^2} \frac{16}{72} \left(
\frac{8}{3}(\frac{\hat{u}}{\hat{t}}+ \frac{\hat{t}}{\hat{u}}
+12\frac{\hat{t} \hat{u}}{\hat{s}^2}-6) 
+\frac{28}{3} \kappa'^2 \hat{s} + 
\frac{16}{3} \kappa'^4 \hat{t} \hat{u} \right)
\end{equation}
For gluon-gluon to quark-antiquark we use the above replacing the
coefficient $16/72$ by $16/256$, for either the quark or antiquark
being observed.

For gluon-quark to gluon-quark with the gluon observed 
(or gluon-antiquark to gluon-antiquark) the cross section is
\begin{equation} 
\frac{d\hat{\sigma}}{d\hat{t}} = 
-\frac{\pi \alpha_s^2}{\hat{s}^2} \frac{16}{96} \left(
\frac{8}{3}(\frac{\hat{s}}{\hat{u}}+ \frac{\hat{u}}{\hat{s}}
+12\frac{\hat{u} \hat{s}}{\hat{t}^2}-6) 
+\frac{28}{3} \kappa'^2 \hat{t} + 
\frac{16}{3} \kappa'^4 \hat{u} \hat{s} \right)
\end{equation}
For gluon-quark to gluon-quark with the quark observed
(or gluon-antiquark to gluon-antiquark) we interchange $\hat{t}$ with
$\hat{u}$ in the above equation.

\subsection{Quark Processes}
For quark-antiquark to the same quark-antiquark including t channel
exchange as well as s channel annihilation with their interference, and
observing the quark we have
\begin{eqnarray}
\frac{d\hat{\sigma}}{d\hat{t}} &=& 
\frac{\pi \alpha_s^2}{\hat{s}^2} \frac{16}{36} 
\left(
\frac{(\hat{u}^2+\hat{t}^2)}{\hat{s}^2}
+\frac{(\hat{s}^2+\hat{u}^2)}{\hat{t}^2}
-\frac{2}{3} \frac{\hat{u}^2}{\hat{s} \hat{t}} \right. \nonumber\\
&&\left. +\kappa'^2 
\left(
4 \hat{s} \hat{t} \hat{u} 
(\frac{1}{\hat{s}^2}+\frac{1}{\hat{t}^2})
+\frac{1}{3}\hat{u} 
(9 + \frac{\hat{s}}{\hat{t}} +\frac{\hat{t}}{\hat{s}}) 
\right) \right. \nonumber\\
&&\left. 
+\frac{1}{2}\kappa'^4 \left( (\hat{t}-\hat{u})^2+(\hat{s}-\hat{u})^2
-\frac{1}{3}(\hat{t}-\hat{u})(\hat{s}-\hat{u}) \right) 
\right)
\end{eqnarray}
For the above process observing the antiquark, we interchange
$\hat{t}$ and $\hat{u}$ in the above formula.
For identical quarks scattering to the same indentical quarks, we 
interchange $\hat{s}$ and $\hat{u}$ in the above equation, and add
a factor of 1/2 for the indentical final state quarks.

For a quark-antiquark annihilation to a virtual gluon creating 
a different quark-antiquark pair, as in charm and bottom production,
summing over identical cross-sections for either quark or antiquark
observed (which will later be multiplied by 4 for four other light
quarks being created) we have for the s-channel gluon exchange
\begin{equation}
\frac{d\hat{\sigma}}{d\hat{t}} = 
\frac{\pi \alpha_s^2}{\hat{s}^2} \frac{16}{36}
\left(\frac{(\hat{u}^2 +\hat{t}^2)}{\hat{s}^2}
+4 \kappa'^2 \hat{s} \hat{t} \hat{u} \frac{1}{\hat{s}^2}
+ \frac{1}{2}\kappa'^4 (\hat{t}-\hat{u})^2 \right)
\end{equation}
For $q+\bar{q'} \to q+\bar{q'}$ observing the final quark we have
for the t-channel gluon exchange
\begin{equation}
\frac{d\hat{\sigma}}{d\hat{t}} = 
\frac{\pi \alpha_s^2}{\hat{s}^2} \frac{16}{36}
\left(\frac{(\hat{s}^2 + \hat{u}^2)}{\hat{t}^2}
+4\kappa'^2 \hat{s} \hat{t} \hat{u} \frac{1}{\hat{t}^2}
+\frac{1}{2}\kappa'^4 (\hat{s}-\hat{u})^2 \right)
\end{equation}
For observing the final antiquark, we use the above equation with
$\hat{t}$ and $\hat{u}$ exchanged.  For $q+q'$ to $q+q'$ observing the
quark we use the above formula and interchange $\hat{s}$ and
$\hat{u}$ in the large parenthesis.  For $q+q'$ to $q+q'$ observing
$q'$, we interchange $\hat{t}$ and $\hat{u}$ in the large
parenthesis of the above formula.

\section{Comparison With Possible Anomalous Magnetic Moments
In \boldmath{$Z$} Couplings}

The possibility of anomalous magnetic moments of quarks appearing at
LEP and SLD at the $Z$ peak and their forward-backward asymmetry has been 
considered\cite{shawsilverman,rizzolep}.  Here we just get 
values to indicate the order of magnitude, refering the reader
to the more careful and complete analyses by T.G. Rizzo\cite{rizzolep},
which gives $-0.012 \leq \kappa^e_b \leq -0.002$ at 95\% CL, updated to
Moriond '96 data\cite{rizzoprivate}.  With an anomalous electric dipole
moment as well, it allows positive $\kappa^e_b \leq 0.025$.
Using the integrated cross-section\cite{shawsilverman,rizzolep}
for $b$ production at the $Z$ peak
we find with $m_b = 4.5$ GeV (and with $F'_2 = \kappa^e_b$)
 
\begin{eqnarray}
\sigma &\propto& (v^2_b + a^2_b + 3v_b \kappa_b^e
+m^2_Z \kappa_b^{e2} /(8 m^2_b) ) \\
\sigma &\propto& (0.365 - 1.03 \kappa^e_b + 51.1 \kappa^{e2}_b).
\end{eqnarray}

Letting the $\kappa^e_b$ terms account for the discrepancy\cite{LEP}
$\Delta R_b/R_b = 0.026$
gives two solutions, $\kappa_b^e = 0.027$, and $\kappa_b^e = -0.0069$, 
corresponding to $\kappa^{'e}_b = 3.0 $ TeV$^{-1}$ and
$\kappa^{'e}_b = -0.77 $ TeV$^{-1}$.

$A^b_{\text{FB}}$ isolates the $\cos{\theta}$ term in the 
$Z$ cross-section which can be partly 
due to the anomalous magnetic moment\cite{shawsilverman,rizzolep}
($A^{\text{SM}}_b \equiv 2 v_b a_b / (v^2_b + a^2_b)$)

\begin{equation}
A_{\text{FB}}^b = \frac{3}{4} A_b^{\text{SM}} A_e 
(1+\kappa^e_b/g_V)/(1 + \Delta R_b/R_b)
\end{equation}
where the inverse power term has been taken here to match $1+\Delta R_b/R_b$.
Using the SM value for $A^{\text{SM}}_b$, the positive $\kappa^e_b =
0.027$ gives a $3.1\sigma$ discrepancy with\cite{LEP} $A_{\text{FB}}^b
= 0.1002 \pm 0.0028$, and is eliminated.  The negative $\kappa^e_b = -
0.0069$ ($\kappa^{'e}_b = -0.77 $ TeV$^{-1}$) value only gives a $0.4
\sigma$ deviation and is consistent.  The error on $A_b =
A^{\text{SM}}_b (1+\kappa^3_b/v_b)/(1+\Delta R_b/R_b)$ is about 6\%,
and it is 11\% below theory.  The correction of the anomalous moment
only lowers theory by $0.5$\%.

Our calculated CDF jet cross-sections do not depend on the sign of
$\kappa^{'g}$, and for the case of only the $b$ quark having the anomalous
chromomagnetic moment $|\kappa^{'g}_b| = 3.5 \pm 1 $ TeV$^{-1}$ is not
inconsistent with the $\Delta R_b$ anomalous magnetic moment
since there will be a numerical factor depending on how
the electric and color charges are distributed among the internal
constituents with different masses.

For the case where all quarks except top are given the anomalous
chromomagnetic moment, allowing an equal anomalous magnetic moment for
each, we find changes in $\Delta \Gamma_{\text{had}}/
\Gamma_{\text{had}}$ of $0.0017$, $0.0005$, and $-0.0005$,
corresponding to the $\kappa'= 1.3$, $1.0$, and $0.7$ TeV$^{-1}$
cases, respectively.
This is at most a one $\sigma$ discrepancy since the fractional
error\cite{LEP} on $\Delta \Gamma_{\text{had}}/\Gamma_{\text{had}}$ is
$\pm 0.0019$.

\section{Results and Conclusions}
In calculating the jet transverse energy distributions, we have used
the parton distribution functions MRSA$'$\cite{mrs}.  Since we are
taking ratios of anomalous moment contributions to strict QCD, both in
leading order, most variance with PDFs drops out, only being about 2\%
at $E_T = 450$ GeV among the MRS set of PDFs, or between MRSA$'$ and
CTEQ3M.  Assuming all quarks except the top have the same anomalous
chromomagnetic moment, we find in Fig.~\ref{fitsall} that the range
$|\kappa'| = 1.0 \pm 0.3 $ TeV$^{-1}$ will fit the CDF
data\cite{CDF,CDFb}, and be in the allowed systematic error range of
the D0 data\cite{D0}.  The quadratic and quartic terms in $\kappa'$
give a natural fit to the possible curvature in the data, and indicate
why the corrections do not show up significantly until $E_T \ge 200$
GeV.  It is not inconsistent with an equal anomalous magnetic moment
being present for the quarks at the $Z$ peak.

If the charm or bottom quark alone among the lighter quarks possesses
a sizeable anomalous chromomagnetic moment, we find in
Fig.~\ref{fitsbc} the range $|\kappa'_{b,c}| = 3.5 \pm 1$ TeV$^{-1}$ will fit
the CDF data, and be in the allowed range of the D0 data.  It is not
inconsistent with a comparable anomalous magnetic moment
explanation\cite{rizzolep} of $R_b$.

\acknowledgements The author thanks Prof. Myron Bander for help with
the conceptual framework and numerous helpful discussions.  We also
thank W. Giele for supplying the structure function program.  The
author thanks the SLAC theory group for their hospitality and
acknowledges discussions with S. Drell, S. Brodsky, L. Dixon, and T.G.
Rizzo.  This research was supported in part by the U.S. Department of
Energy under Contract No. DE-FG03-91ER40679.

\begin {figure}
\caption{The ratio of the anomalous chromomagnetic moment contribution to
the lowest order QCD jet distribution in transverse energy.  The
dashed, solid, and dotted curves are for $|\kappa'| =1.3$, $1.0$,
and $0.7$ TeV$^{-1}$, respectively.  The crossed data are from CDF
Run 1a, and the circle data from Run 1b (preliminary).}
\label{fitsall}
\end{figure}
 
\begin{figure}
\caption{The ratio of the anomalous chromomagnetic moment contribution
of either a bottom or charm quark-antiquark pair created to the 
lowest order QCD jet distribution in transverse energy.  The dashed,
solid, and dotted curves are for $|\kappa'_{b,c}|= 4.5$, $3.5$, and 
$2.5$ TeV$^{-1}$,
respectively.  The data are as in Fig.~1.}
\label{fitsbc}
\end{figure}

\end{document}